\documentclass[11pt]{article}

\usepackage[dvips]{epsfig}

\usepackage{fullpage}
\usepackage{amsmath}
\usepackage{amssymb}
\usepackage{amsthm}
\usepackage{algorithm}
\usepackage{algorithmic}
\usepackage{enumerate}
\usepackage{float}

\newtheorem{Theorem}{Theorem}
\newtheorem{Lemma}{Lemma}
\newtheorem{Corollary}{Corollary}
\newtheorem{Fact}{Fact}

\newtheoremstyle{definition}{}{}
     {\rmfamily}%         Body font
     {}%         Indent amount (empty = no indent, \parindent = para indent)
     {\bfseries}% Thm head font
     {.}%        Punctuation after thm head
     { }%     Space after thm head (\newline = linebreak)
     {}%         Thm head spec
\theoremstyle{definition}
\newtheorem{Definition}{Definition}

\newenvironment{Proof}[0]{\begin{proof}}{\end{proof}}

\newcommand{\Nm}{\mathbb{N}}

\newcommand{\Zm}{\mathbb{Z}}

\renewcommand{\epsilon}{\varepsilon}

\newcommand{\toset}{\rightarrow}

\newcommand{\DQC}{\mathsf{DQC}}
\newcommand{\RQC}{\mathsf{RQC}}
\newcommand{\QQC}{\mathsf{QQC}}

\renewcommand{\phi}{\varphi}

\newcommand{\boundary}[1][]{\partial\,{#1}}
\newcommand{\boundaryn}[2][]{\partial_{#1}\,{#2}}

\floatstyle{plain}
\restylefloat{figure}

\begin{document}

\title{On the black-box complexity of Sperner's~Lemma}

\author{
Katalin Friedl%
\thanks{%
{Budapest University of Technology and Economics,}
{H-1521 Budapest, P.O.Box 91., Hungary.}
Email: {\tt fri\hspace{-20cm}.\hspace{19.8cm}edl@cs.bme.hu}.
The research was supported by the EU 5th framework program
RESQ IST-2001-37559 and Centre of Excellence ICAI-CT-2000-70025,
and OTKA grants T42559 and T46234.
}
\and
G\'abor Ivanyos%
\thanks{%
{Computer and Automation Research Institute,
Hungarian Academy of Sciences.}
{H-1518 Budapest, P.O. Box 63., Hungary.}
Email: {\tt Gabor.Iva\hspace{-20cm}.\hspace{19.8cm}nyos@sztaki.hu}.
The research was supported by the EU 5th framework program RESQ IST-2001-37559
and Centre of Excellence ICAI-CT-2000-70025,
and OTKA grants T42706 and T42481.
}
\and
Miklos Santha
\thanks{%
{CNRS--LRI, UMR 8623},
{b\^atiment 490,
Universit\'e Paris XI,
91405 Orsay, France.}
Email: {\tt san\hspace{-20cm}.\hspace{19.8cm}tha@lri.fr}.
The research was supported by the EU
5th framework program RESQ IST-2001-37559, and by the ACI CR 2002-40
and ACI SI 2003-24 grants of the French Research Ministry.
}
\and
Yves F. Verhoeven
\thanks{%
{LRI, UMR 8623,
b\^atiment 490,
Univ. Paris XI,
91405 Orsay}
and
{ENST,  46 rue Barrault,
75013 Paris, France.}
Email: {\tt yves.verho\hspace{-20cm}.\hspace{19.8cm}even@normalesup.org}.
The research was supported by the EU 5th framework program RESQ IST-2001-37559,
and by the ACI CR 2002-40 and ACI SI 2003-24 grants of the
French Research Ministry.
}
}
\date{}

\maketitle
\vspace{-3mm}
\begin{abstract}
We present several results on the complexity of various forms of Sper\-ner's Lemma
in the black-box model of computing. We give a deterministic algorithm for Sperner problems
over pseudo-manifolds of arbitrary dimension. The query complexity of our algorithm is %essentially
linear  in the separation number of the skeleton graph of the manifold and the size of its boundary.
As a corollary we get an $O(\sqrt{n})$ deterministic query algorithm for the black-box version
of the problem {\bf 2D-SPERNER}, a well studied member of Papadimitriou's complexity
class PPAD.
This upper bound matches the $\Omega(\sqrt{n})$
deterministic
lower bound of Crescenzi and Silvestri.
The tightness of this bound was not known before.
In another result
we prove for the same problem an
$\Omega(\sqrt[4]{n})$ lower bound for its probabilistic, and an
$\Omega(\sqrt[8]{n})$ lower bound for its 
quantum query complexity, showing that all these measures are
polynomially related. 
\end{abstract}

\paragraph{Classification:}
computational and structural complexity,
quantum computation and information.

\section{Introduction}
\label{section:introduction}
Papadimitriou defined in~\cite{pap,papadimitriou} the 
complexity classes PPA, PPAD, and PSK 
in order to classify total search problems
which have always a solution.The class PSK was renamed
PPADS in~\cite{Beame}.
These classes can be characterized by some underlying combinatorial 
principles. The class Polynomial Parity Argument~(PPA) is the class of NP search 
problems, where the existence of the solution is guaranteed by the fact that
in every finite graph the number of vertices with odd degree is even.
The class PPAD is the directed version of PPA, and its basic  search
problem  is the following: in a directed graph, 
where the in-degree and the out-degree of every vertex is at most one,
given a source, find
another source or a sink. In the class PPADS the basic search problem is more restricted than
in PPAD: given a source, find a sink.

These classes are in fact subfamilies of TFNP,
the family of all total NP-search problems,
introduced by Megiddo and Papadimitriou~\cite{Megiddo_Papadimitriou}.
Other important subclasses of TFNP are Polynomial
Pigeonhole Principle (PPP) and Polynomial Local Search (PLS).
The elements of PPP are problems which by
their combinatorial nature obey the pigeonhole principle and
therefore have a solution. In a PLS problem, one is looking for
a local optimum for a particular objective function, in
some neighborhood structure.
All these classes are interesting because
they contain search problems not known to be
solvable in polynomial time, but which are also somewhat easy
in the sense that they can not be NP-hard unless NP = co-NP.

Another point that makes the parity argument classes interesting is that there are
several natural problems from different branches of mathematics that
belong to them.
For example, in a graph with odd degrees, when a
Hamiltonian path is given, a theorem of Smith~\cite{Thomason}
ensures that there is another Hamiltonian path. 
It turns out that finding this second path belongs to the class 
PPA~\cite{papadimitriou}. A search problem coming from a modulo 2 
version of Chevalley's theorem~\cite{papadimitriou} from number theory is also in PPA.
Complete problems in PPAD
are the search versions of Brouwer's fixed point theorem, Kakutani's fixed point theorem, Borsuk-Ulam
theorem, and Nash equilibrium (see~\cite{papadimitriou}).

The classical Sperner's Lemma~\cite{Sperner} states that in a triangle with a regular triangulation
whose vertices are labeled with three colors, there is always a trichromatic triangle.
This lemma is of special interest since some customary proofs for the above topological
fixed point theorems rely on its combinatorial content. However, it is unknown whether
the corresponding search problem, that Papadimitriou~\cite{papadimitriou} calls {\bf 2D-SPERNER},
is complete in PPAD. Variants of Sperner's Lemma also give rise to other problems in 
the parity argument classes.
Papadimitriou~\cite{papadimitriou} has proven that a 3-dimensional
analogue of {\bf 2D-SPERNER} is in fact complete in PPAD.

The study of query complexities of the 
black-box versions of several problems in TFNP is an active field of research.
Several recent results point into the direction that quantum algorithms
can give only a limited speedup over deterministic ones in this framework.
The collision lower bound of Aaronson~\cite{Aar2} and
Shi~\cite{Shi} about PPP, and the recent result of Santha and Szegedy~\cite{Miklos_Mario}
on PLS imply that the respective deterministic and quantum complexities are polynomially related.
As a consequence, if an efficient quantum algorithm exists for a problem in these classes,
it must exploit its specific structure.
In a related issue, Buresh-Oppenheim and Morioka~\cite{Morioka} have obtained relative
separation results among PLS and the polynomial parity argument classes.

\section{Results}
\label{section:results}
A {\em black-box problem} is a relation
$R \subseteq S \times T$ where  $T$ is a finite set and
$S\subseteq \Sigma^n$ for some finite
set $\Sigma$.
The oracle input is a function $x \in S$, hidden by
a black-box, such that $x_i$, for $i\in \{1,\ldots,n\}$
can be accessed
via a query parameterized by $i$. The output of the problem is
some $y\in T$ such that $(x,y) \in R$.
A special case is the
{\em functional oracle problem}  when the relation is  given by a function
$A : S \rightarrow T$, the (unique) output is then $A(x)$.
We say that $A$ is {\em total} if $S = \Sigma^n$.

In the query model of computation each query adds one to 
the complexity of the algorithm, but
all other computations are free.
The state of the computation is represented by three
registers, the query register $i \in \{1,\ldots,n\}$, the answer register $a \in \Sigma$, and the 
work register $z$. The computation takes place in the vector space spanned by all
basis states $|i\rangle|a\rangle|z\rangle$.
In the {\em quantum query model} introduced by Beals, Buhrman, Cleve, Mosca and de Wolf~\cite{BBC}
the state of the computation is a complex
combination of all basis states which has unit length in the norm $l_2$.
In the randomized model it is a non-negative real combination of unit length
in the norm $l_1$, and in the deterministic model it is always one of the basis states.

The query operation $O_x$ maps the 
basis state
$|i\rangle|a\rangle|z\rangle$ 
into the state $|i\rangle|(a+x_i) \bmod |\Sigma|\rangle|z\rangle$
(here we identify $\Sigma$ with the residue classes $\bmod |\Sigma|$).
Non-query operations are independent of $x$.
A {\em $k$-query algorithm} is a sequence of $(k+1)$ operations
$(U_0, U_1, \ldots , U_k)$ where $U_i$ is unitary in the quantum 
and stochastic in the randomized model, and it is a permutation in the deterministic case.
Initially the state of the computation is set to some
fixed value $|0\rangle|0\rangle|0\rangle$, and then the sequence of operations
$U_0, O_x, U_1, O_x, \ldots, U_{k-1}, O_x, U_k$ is applied.
A quantum or randomized
algorithm computes (with two-sided error) $R$ if the observation of the appropriate last bits of the work register
yield some $y \in T$ such that $(x,y) \in R$ with probability 
at least $2/3$. 
Then $\QQC(R)$ 
(resp. $\RQC(R)$) is the smallest $k$ for which there exists
a $k$-query quantum (resp. randomized) algorithm which 
computes $R$. In the case of deterministic algorithms
of course exact computation
is required, and the deterministic query complexity $\DQC(R)$
is defined then analogously.
We have 
$
\DQC(R) \ge  \RQC(R)  \ge
\QQC(R)$.

Beals et al.~\cite{BBC} have shown that in the case of total
functional oracle problems
the deterministic and quantum
complexities are polynomially related, and the gap is at most a degree 6 polynomial.
For several partial functional problems
exponential quantum speedups are known~\cite{DJ,Sim}.

In this paper we will give several results about Sperner problems in the black-box
framework. In Section~\ref{section:algorithms}, we will prove that the deterministic query complexity of {\bf REGULAR 2-SPM}, the
black-box version of {\bf 2D-SPERNER} is $O(\sqrt{n})$.
This matches the deterministic $\Omega(\sqrt{n})$ lower bound of Crescenzi and Silvestri~\cite{Crescenzi}. 
The tightness of this bound was not known before.
In fact, this result is the corollary of a general algorithm
that solves the Sperner problems over pseudo-manifolds
of arbitrary dimension. The complexity analysis of the algorithm will be expressed in {\bf Theorem~\ref{th:abstract_complexity}}
in two combinatorial parameters of the pseudo-manifold: the size of its boundary and the separation number of its skeleton graph. 
In Section~\ref{section:bounds}, we show that quantum, probabilistic, and deterministic query complexities of
{\bf REGULAR 2-SPM} are polynomially related. More precisely, in {\bf Theorem~\ref{th:blackbox}} we will prove
that its randomized complexity is $\Omega(\sqrt[4]{n})$ and that its quantum complexity is $\Omega(\sqrt[8]{n})$.
This result is analogous to the polynomial relations obtained for the respective query complexities of PPP and PLS.

\section{Mathematical background on simplicial complexes}
\label{sections:maths}
For an undirected graph $G=(V,E)$, and for a subset $V'\subseteq V$ of the vertices,
we denote by $G[V']$ the induced subgraph of $G$ by $V'$.
A graph $G''=(V'',E'')$ is a subgraph of $G$, in notation $G''\subseteq G$, if $V''\subseteq V$
and $E''\subseteq E$. 
The ring $\Zm/(2)$ denotes the ring with $2$ elements.

\begin{Definition}[Simplicial complex]
A {\em simplicial complex} $K$ is a non-empty collection of subsets of a finite set $U$,
such that whenever $S\in K$ then $S'\in K$
for every $S'\subseteq S$. An element $S$ of $K$ of cardinality $d+1$
is called a {\em $d$-simplex}.
A $d'$-simplex $S'\subseteq S$ is called
a {\em $d'$-face of $S$}.
We denote by $K_d$ the set of $d$-simplices of $K$.
An {\em elementary $d$-complex} is
a simplicial complex that contains exactly one $d$-simplex and its subsets.
The {\em dimension} of $K$, denoted by $\dim(K)$, is the largest $d$ such that $K$
contains a $d$-simplex.

The elements of $K_0$ are called the {\em vertices} of $K$, and the
elements of $K_1$ are called the {\em edges} of $K$.
The {\em skeleton graph} $G_K=(V_K,E_K)$ is the
graph whose vertices are the vertices of $K$, and the edges are the edges of $K$.
\end{Definition}

Without loss of generality, we suppose that $U$ consists of integers,
and we identify $\{u\}$ with $u$, for $u\in U$.

\begin{Fact}
\label{fact:complete_graph}
Let $d$ be a positive integer. If $S$ is an elementary $d$-complex, then $G_S$ is the complete graph.
\end{Fact}

\begin{Definition}[Oriented Simplex]
For every positive integer $n$, we define an equivalence relation $\equiv_n$
over $\Zm^n$, by $a \equiv_n b$ if there
exists an even permutation $\sigma$ such that $\sigma\cdot a=b$.
For every $a\in\Zm^n$ we denote by $[a]_{\equiv_n}$ the equivalence
class of $a$ for $\equiv_n$.
The two equivalence classes of the orderings of the $0$-faces of
a simplex are called its {\em orientations}. An {\em oriented simplex}
is a pair formed of a simplex and one of its orientations.
\end{Definition}

For an oriented $d$-simplex $(S,[\tau]_{\equiv_{d+1}})$, where $\tau$ is an
ordering of the $0$-faces of $S$, and a permutation $\sigma$
over $\{1,\ldots,d+1\}$, we denote by $\sigma\cdot (S,[\tau]_{\equiv_{d+1}})$ the
oriented $d$-simplex $(S,[\sigma\cdot \tau]_{\equiv_{d+1}})$.
For every integer $d$, and every simplicial complex $K$ whose simplices have
been oriented, we denote by $K_d$ the set of oriented $d$-simplices of $K$.
From now on, $S$ may denote an oriented or a non-oriented simplex.
When $S$ is an oriented simplex, $\bar{S}$ will denote the same simplex
with the opposite orientation. We also define $S^{(i)}$ to be $S$ if $i$ is even,
and to be $\bar{S}$ if $i$ is odd.
We will often specify an oriented simplex by an ordering
of its $0$-faces.

\begin{Definition}
Let $S=(v_0,\ldots,v_d)$ be an oriented $d$-simplex.
For every $0\leq i\leq d$, for every $(d-1)$-face  $\{v_0,\ldots,v_{i-1},v_{i+1},\ldots,v_d\}$ of $S$,
the {\em induced orientation} is the oriented $(d-1)$-simplex $(v_0,\ldots,v_{i-1},v_{i+1},\ldots,v_d)^{(i)}$.
\end{Definition}

\begin{Definition}
Let $K$ be a simplicial complex whose simplices have been oriented, and let $R$ be a ring.
We define $C_d(K;R)$ as the submodule of the free $R$-module over the
$d$-simplices of $K$ with both possible orientations, whose elements are of the form
$\sum_{S\in K_d} (c_S\cdot S+c_{\bar{S}}\cdot\bar{S})$, with $c_S\in R$,
satisfying the relation $c_S=- c_{\bar{S}}$.
The elements of $C_d(K;R)$ are called {\em $d$-chains}.
For every oriented simplex $S$ of $K$, we denote by $\langle S\rangle$ the
element $S-\bar{S}$ of $C_d(K;R)$.

Let $S$ be an oriented $d$-simplex $(v_0,v_1,\ldots v_d)$ of $K$. The
{\em algebraic boundary} of
$\langle S\rangle$, denoted by $\boundaryn[d]{\langle S\rangle}$,
is the $(d-1)$-chain of $C_{d-1}(K;R)$ defined as
$\boundaryn[d]{\langle S\rangle}=\sum_{i=0}^d (-1)^i \langle(v_0,\ldots,v_{i-1},v_{i+1},\ldots, v_d)\rangle$.
\end{Definition}

Since $\boundaryn[d]{\langle S\rangle}=-\boundaryn[d]{\langle \bar{S}\rangle}$,
the operator $\partial_d$ has been correctly defined on a basis of
$C_d(K;R)$ and can therefore be uniquely extended into a homomorphism
$\partial_d:C_d(K;R)\toset C_{d-1}(K;R)$.
The proof of the next Lemma is straightforward.

\begin{Lemma}
\label{lemma:boundary}
Let $S$ be an oriented $d$-simplex of a simplicial complex $K$. Denote by $F_S$ the set
of $(d-1)$-faces of $S$, and for every $S'\in F_S$ by $\tau_{S'}^S$
the induced orientation on $S'$.
Then $\boundaryn[d]{\langle S\rangle}=\sum_{S'\in F_S} \langle(S',\tau_{S'}^S)\rangle$.
\end{Lemma}

Following an early version of a paper of Bloch~\cite{Bloch},
in the next definition we generalize the notion of pseudo-manifold, without
the usual requirements of connectivity and pure dimensionality.
\begin{Definition}
\label{def:pseudomanifold}
A simplicial complex $\mathcal{M}$ is a {\em pseudo $d$-manifold}, for a positive integer
$d$, if
\begin{enumerate}[(i)]
\item $\mathcal{M}$ is a union of elementary $d$-complexes,
\item every $(d-1)$-simplex in $\mathcal{M}$ is a $(d-1)$-face of at most two $d$-simplices
  of $\mathcal{M}$.
\end{enumerate}
The {\em boundary of $\mathcal{M}$} is the set of elementary $(d-1)$-complexes in $\mathcal{M}$ that belong
exactly to one $d$-simplex of $\mathcal{M}$. We denote it by $\boundary[\mathcal{M}]$.
A pseudo $d$-manifold $\mathcal{M}$ is said to be {\em orientable} if it is possible
to assign an orientation to each $d$-simplex of $\mathcal{M}$, such that for all
$(d-1)$-simplex of $\mathcal{M}$ that is not on its boundary the orientations
induced by the two $d$-simplices to which it belongs are opposite.
Such a choice of orientations for all the $d$-simplices of $\mathcal{M}$
makes $\mathcal{M}$ {\em oriented}.
\end{Definition}

If the $d$-simplices of $\mathcal{M}$ are oriented, then there is a natural orientation
of the $(d-1)$-simplices of $\boundary[\mathcal{M}]$, where each $(d-1)$-simplex has the orientation
induced by the oriented $d$-simplex of which it is a $(d-1)$-face.
Notice that if $\mathcal{M}$ is a pseudo $d$-manifold, then
$\boundary[M]$ need not be a pseudo $(d-1)$-manifold.
From now, all the simplicial complexes will be pseudo-manifolds.
Observe that if $R=\Zm/(2)$, then for any oriented $d$-simplex $S$, we have $\langle S\rangle=\langle\bar{S}\rangle$.

\begin{Definition}
Given a simplicial complex $K$ of dimension $d$, the {\em standard $d$-chain $\widehat{K}$ of $K$}
will be defined depending on whether $K$ is oriented as follows:
\begin{itemize}
\item if $K$ is non-oriented, then $\widehat{K}=
  \sum_{S\in K_d} \langle (S,\tau_S)\rangle\in C_d(K,\Zm/(2))$,
  for an arbitrary choice of orientations $\tau_S$ of the $d$-simplices $S$ in $K$,
\item if $K$ is oriented, then $\widehat{K}=
  \sum_{S\in K_d} \langle (S,\tau_S)\rangle\in C_d(K,\Zm)$
  where $\tau_S$ is the orientation of $S$ in $K$.
\end{itemize}
\end{Definition}

\begin{Fact}
\label{fact:pseudo_boundary}
Let $d$ be an integer, and let $\mathcal{M}$ be a pseudo $d$-manifold. Then,
\begin{enumerate}[1.]
\item if $\mathcal{M}$ is not oriented the equality $\widehat{\boundary[\mathcal{M}]}=\boundaryn[d]{\widehat{\mathcal{M}}}$
  holds in $C_{d-1}(\boundary[\mathcal{M}],\Zm/(2))$,
\item and if $\mathcal{M}$ is oriented the equality $\widehat{\boundary[\mathcal{M}]}=\boundaryn[d]{\widehat{\mathcal{M}}}$
  holds in $C_{d-1}(\boundary[\mathcal{M}],\Zm)$.
\end{enumerate}
\end{Fact}

\begin{Proof}
For every $d$-simplex $S$ in $\mathcal{M}$, denote by $F_S$ the
set of $(d-1)$-faces of $S$, and for every $S'\in F_S$ denote by $\tau_{S'}^S$ the induced orientation of $S'$ from $S$.
From Lemma~\ref{lemma:boundary}, we can write
$$\boundaryn[d]{\widehat{\mathcal{M}}}=\sum_{S\in\mathcal{M}_d} \boundaryn[d]{\langle(S,\tau_S)\rangle}=
  \sum_{S\in\mathcal{M}_d}\sum_{S'\in F_S} \langle(S',\tau_{S'}^S)\rangle.$$
From the definition of a pseudo $d$-manifold, we know that in the last sum each $(d-1)$-simplex that
is not on the boundary of $\mathcal{M}$ appears exactly twice.
If $\mathcal{M}$ is not oriented, then as the base ring is $\Zm/(2)$, the only $(d-1)$-simplices
that remain in the sum are those that are in $\boundary[\mathcal{M}]$.
If $\mathcal{M}$ is oriented, then from the definition of the orientability, it follows that each $(d-1)$-simplex
that appears in two $d$-simplices of $\mathcal{M}$ appears in the sum once with each orientation.
As for any oriented simplex $S$ the equality $\langle S\rangle+\langle\bar{S}\rangle=0$ holds,
the only terms that do not cancel are the oriented $(d-1)$-simplices of the boundary.
These $(d-1)$-simplices appear with the correct orientation.
\end{Proof}

\section{Sperner Problems}
\label{section:sperner}

\begin{Definition}
Let $K$ be a simplicial complex. A {\em labeling of $K$} is a mapping $\ell$ of the
vertices of $K$ into the set $\{0,\ldots,\dim(K)\}$.
If a simplex $S$ of $K$ is labeled with all possible labels, then we say that {\em $S$ is fully labeled}.
\end{Definition}

A labeling $\ell$ naturally maps every oriented $d$-simplex $S=(v_0,\ldots,v_d)$ to 
the equivalence class $\ell(S)=[(\ell(v_0),\ldots,\ell(v_d))]_{\equiv_{d+1}}$.

\begin{Definition}
Given a labeling $\ell$ of a simplicial complex $K$, and an integer $0\leq d\leq\dim(K)$,
we define the {\em $d$-dimensional flow} $N_d[\langle S\rangle]$ by
$$ N_d[\langle S\rangle]=\begin{cases}
1 & \text{if } \ell(S)=[(0,1,2\ldots, d)]_{\equiv_{d+1}}, \\ 
-1 & \text{if } \ell(S)=[(1,0,2,\ldots,d)]_{\equiv_{d+1}}, \\
0 & \text{otherwise}, 
\end{cases}$$
and then extend it by linearity into a homomorphism $N_d:C_d(K;R)\toset R$.
\end{Definition}

Sperner's Lemma~\cite{Sperner} has been generalized in several ways. The following
statement from~\cite{Taylor} is also a straightforward consequence of results of~\cite{KyFan}.
\begin{Theorem}[Sperner's Lemma~\cite{Sperner,KyFan,Taylor}]
\label{th:sperner_general}
Let $K$ be a simplicial complex of dimension $d$, let $\ell$ be a labeling of $K$, and let $R$ be a ring.
For an element $C$ of $C_d(K;R)$, we have $N_d[C]=(-1)^d N_{d-1}[\boundaryn[d]{C}]$. 
\end{Theorem}

Using Fact~\ref{fact:pseudo_boundary}, 
we translate Theorem~\ref{th:sperner_general} into terms of pseudo-manifolds.
\begin{Theorem}[Sperner's Lemma on pseudo-manifolds]
\label{th:sperner_pseudomanifold}
Let $d$ be an integer, let $\mathcal{M}$ be a pseudo $d$-manifold, and let
$\ell$ be a labeling of $\mathcal{M}$. Then 
$ N_d[\widehat{\mathcal{M}}]=(-1)^d N_{d-1}[\widehat{\boundaryn{\mathcal{M}}}]$
where
$$
\begin{cases}
\widehat{\mathcal{M}}\in C_d(\mathcal{M},\Zm/(2)),\, \widehat{\boundaryn{\mathcal{M}}} \in C_{d-1}(\boundaryn{\mathcal{M}},\Zm/(2)),
  & \text{ if } \mathcal{M} \text{ is not oriented,} \\
\widehat{\mathcal{M}}\in C_d(\mathcal{M},\Zm),\, \widehat{\boundaryn{\mathcal{M}}} \in C_{d-1}(\boundaryn{\mathcal{M}},\Zm),
  & \text{ if } \mathcal{M} \text{ is oriented.}
\end{cases}
$$
\end{Theorem}

This version of Sperner's lemma can be viewed, from a physicist's point of view, as a result equivalent to a
global conservation law of a flow.
If there is a source for the flow and the space is bounded then there must be a sink for that flow.
More concretely, the lines of flow can be drawn over $d$-simplices, that goes from
one $d$-simplex to another if they share a $(d-1)$-face that has all possible labels in $\{0,\ldots,d-1\}$.
The sources and sinks of the flow are the fully labeled $d$-simplices.
The lemma basically says that 
if the amount of flow entering the manifold at the boundary is larger than 
the exiting flow, then there must exist sinks inside.
The local conservation is stated by the fact that if there is an ingoing edge,
there will not be two outgoing edges, and conversely. Formally, we have the following.

\begin{Fact}
\label{fact:simplex}
Let $(S,\tau_S)$ be an oriented $d$-simplex. Then at most
two of its oriented $(d-1)$-faces have a non-zero image by $N_{d-1}$.
Moreover, if there are exactly two $(d-1)$-faces $(S',\tau_{S'}^S)$ and $(S'',\tau_{S''}^S)$
that have non-zero image by $N_{d-1}$, then
$N_d[\langle (S,\tau_{S})\rangle]=0$ and
$N_{d-1}[\langle (S',\tau_{S'}^S)\rangle]=-N_{d-1}[\langle (S'',\tau_{S''}^S)\rangle]$.
\end{Fact}

This gives a relation between the problem of finding fully labeled $d$-simplices and
the natural complete problems for the parity argument classes. We can consider
an oriented $d$-simplex $(S,\tau_S)$ with $N_d[\langle (S,\tau_S)\rangle]=1$
as a source for the flow, and $(S',\tau_{S'})$ with $N_d[\langle (S',\tau_{S'})\rangle]=-1$
as a sink.

We now state the black-box Sperner problems we will consider. 
{
\vskip 2mm
\noindent {\bf Sperner on Pseudo $d$-Manifolds ($d$-SPM)} \\
\begin{tabular}{ll}
  {\it Input:} & a pseudo $d$-manifold $\mathcal{M}$, and $S\in\mathcal{M}_d$. \\ 
  {\it Oracle input:} & a labeling $\ell:\mathcal{M}_0\toset\{0,1,\ldots,d\}$. \\
  {\it Promise:} & one of the two conditions holds: \\
      & \hskip 0.5cm {\it a)} $N_{d-1}[\widehat{\boundary\mathcal{M}}]=1$ in $C_{d-1}(\boundary[\mathcal{M}],\Zm/(2))$, \\
      & \hskip 0.5cm {\it b)} $N_{d-1}[\widehat{\boundary\mathcal{M}}]=0$ in $C_{d-1}(\boundary[\mathcal{M}],\Zm/(2))$ and $N_d[\langle S\rangle]=1$ in $C_d(\mathcal{M},\Zm/(2))$. \\
  {\it Output:} & $S'\in\mathcal{M}_d$ such that $N_d[\langle S'\rangle]=1$, with $S\neq S'$ for case {\em b}. \\
\end{tabular}
\vskip 2mm
}

\vskip 2mm
\noindent {\bf Oriented Sperner on Pseudo $d$-Manifolds ($d$-OSPM)} \\
\begin{tabular}{ll}
  {\it Input:} & an oriented pseudo $d$-manifold $\mathcal{M}$, and $S\in\mathcal{M}_d$. \\ 
  {\it Oracle input:} & a labeling $\ell:\mathcal{M}_0\toset\{0,1,\ldots,d\}$. \\
  {\it Promise:} & one of the two conditions holds: \\
      & \hskip 0.5cm {\it a)} $(-1)^d N_{d-1}[\widehat{\boundary\mathcal{M}}]<0$ in $C_{d-1}(\boundary[\mathcal{M}],\Zm)$, \\
      & \hskip 0.5cm {\it b)} $(-1)^d N_{d-1}[\widehat{\boundary\mathcal{M}}]= 0$ in $C_{d-1}(\boundary[\mathcal{M}],\Zm)$
	 and $N_d[\langle S\rangle]=1$ in $C_d(\mathcal{M},\Zm)$. \\
  {\it Output:} & $S'\in\mathcal{M}_d$ such that $N_d[\langle S'\rangle]=-1$.
\end{tabular}
\vskip 2mm

\noindent We will deal in particular with the following important special case of {\bf $2$-SPM}. Let
$V_m =\{(i,j)\in\Nm^2 \,|\, 0\leq i+j\leq m\}$. Observe that $|V_m|={m+2 \choose 2}$.

{
\vskip 2mm
\noindent {\bf Regular Sperner (REGULAR $2$-SPM)} \\
\begin{tabular}{ll}
  {\it Input:} & $n =  {m+2 \choose 2}$ for some integer $m.$\\
  {\it Oracle input:} & a labeling $\ell:V_m\toset\{0,1,2\}$. \\
  {\it Promise:} & for $0\leq k\leq m$, $\ell(0,k)\neq 1$, $\ell(k,0)\neq 0$, and $\ell(k,m-k)\neq 2$. \\
  {\it Output:} & $p,p'$ and $p''\in V$, such that $p'=p+(\epsilon,0)$, $p''=p+(0,\epsilon)$ for some $\epsilon\in\{-1,1\}$, \\
    &  and $\{\ell(p),\ell(p'),\ell(p'')\}=\{0,1,2\}$.
\end{tabular}
\vskip 2mm
}

\noindent In fact, {\bf REGULAR $2$-SPM} on input $n={m+2 \choose 2}$ is the instance of {\bf $d$-SPM}
on the regular $m$-subdivision of an elementary $2$-simplex.
Theorem~\ref{th:sperner_pseudomanifold} states that 
both $d$-SPM and $d$-OSPM have
always a solution.
The solution is not necessarily unique as it can be easily checked on simple instances. Thus the problems
are not functional oracle problems.

\section{Black-box algorithms for pseudo $d$-manifolds}
\label{section:algorithms}
The purpose of this section is to give a black-box algorithm for {\bf $d$-SPM} and {\bf $d$-OSPM}.
To solve these problems, we adopt a divide and conquer approach.
This kind of approach was successfully used in~\cite{Llewellyn_Tovey_1,Llewellyn_Tovey_2} and~\cite{Miklos_Mario},
to study the query complexity of the oracle version of the Local Search problem.
However, the success of the divide and conquer paradigm for Sperner problems
relies heavily on the use of the very strong statement of Sperner's Lemma that is given in
Theorem~\ref{th:sperner_pseudomanifold}. The usual, simpler version of Sperner's Lemma, like the one
given in~\cite{papadimitriou} does not appear to be strong enough for this purpose. Observe that though the
standard proof of Sperner's Lemma is constructive, it yields only an algorithm of complexity $O(n)$.
In our algorithms the division of the pseudo $d$-manifold $\mathcal{M}$ will be done according to the combinatorial
properties of its skeleton graph. The particular parameter we will need is its {\em iterated separation number}
that we introduce now for general graphs.

\begin{Definition}
Let $G=(V,E)$ be a graph. If $A$ and $C$ are
subsets of $V$ such that
$V=A\cup C$, and that there is no edge between $A\setminus C$ and $C\setminus A$,
then $(A,C)$ is said to be a {\em separation} of the graph $G$, in notation
$(A,C)\prec G$. The set $A\cap C$ is called a {\em separator} of the graph $G$. 

The {\em iterated separation number} is defined by induction on the
size of the
graph $G$ by
$s(G)=
\min_{(A,C)\prec G}\left\{|A\cap C|+\max(s(G[A\setminus C]),s(G[C\setminus A]))\right\}$.
A pair $(A,C)\prec G$ such that $s(G)=|A\cap C|+\max(s(G[A\setminus C]),s(G[C\setminus A]))$ is called a {\em
best separation} of $G$.
\end{Definition}

The iterated separation number of a graph is equal to the {\it value of the
separation game} on the graph $G$, which was introduced in~\cite{Llewellyn_Tovey_1}.
In that article, that value was defined as the gain of a player in a certain game.
Notice, also, that the iterated separation number is at most $\log |V|$ times
the {\em separation number} as defined in~\cite{Miklos_Mario}.
Before giving the algorithms, and their analyses, we still need a few observations.

\begin{Lemma}
\label{lemma:splitting}
Let $\mathcal{A}$ and $\mathcal{B}$ be two pseudo $d$-manifolds, such that $\mathcal{A}\cup\mathcal{B}$
is also a pseudo $d$-manifold. Let $\ell$ be a labeling of $\mathcal{A}\cup\mathcal{B}$.
If $\mathcal{A}$ and $\mathcal{B}$ have no $d$-simplex in their intersection, then
$ N_d[\widehat{\mathcal{A}\cup\mathcal{B}}]=N_d[\widehat{\mathcal{A}}]+N_d[\widehat{\mathcal{B}}]$.
\end{Lemma}

\begin{Lemma}
Let $\mathcal{M}$ be a pseudo $d$-manifold, and 
$\mathcal{M}'$ be a union of elementary $d$-complexes such that
 $\mathcal{M}'\subseteq\mathcal{M}$.
Then $\mathcal{M}'$ is a pseudo $d$-manifold.
\end{Lemma}

\begin{Theorem}
\label{th:splitting}
Let $\mathcal{M}$ be a pseudo $d$-manifold, $H$ a subset of $\mathcal{M}_0$, and $\ell$ be a labeling
of the vertices of $\mathcal{M}$.
Let $(A,C)\prec G_\mathcal{M}[\mathcal{M}_0\setminus H]$, $B=H\cup(A\cap C)$, 
and
$M'=A\setminus C$ and $M''=C\setminus A$. Denote by $\mathcal{B}$ the set of elementary
$d$-complexes of $\mathcal{M}$ whose vertices are all in $B$, and by
$\mathcal{M}'$~({\em resp.} $\mathcal{M}''$) the set of elementary $d$-complexes
of which at least one of the vertices belongs to $M'$~({\em resp.} $M''$).
Denote also by $\mathcal{B}'$ the set of elementary $(d-1)$-complexes of
$\mathcal{M}$ whose vertices are all in $B$. Then,
\vspace{-2mm}
\begin{enumerate}[(i)]
\item $\mathcal{B}$, $\mathcal{M}'$, $\mathcal{M}''$ and $\mathcal{M}'\cup\mathcal{M}''$ are pseudo $d$-manifolds,
\item if $H\neq \mathcal{M}_0$ then $\mathcal{B}$, $\mathcal{M}'$ and $\mathcal{M}''$ are proper subsets of $\mathcal{M}$,
\item $ N_d[\widehat{\mathcal{M}}]=N_d[\widehat{\mathcal{B}}]+N_d[\widehat{\mathcal{M}'}]+N_d[\widehat{\mathcal{M}''}]$,
\item the inclusions $\boundary[\mathcal{M}']\subseteq (\boundary[\mathcal{M}])\cup\mathcal{B}'$
       and $\boundary[\mathcal{M}'']\subseteq (\boundary[\mathcal{M}])\cup\mathcal{B}'$ hold,
\end{enumerate}
\end{Theorem}

\begin{Proof}
Clearly, the complexes $\mathcal{B}$, $\mathcal{M}'$, $\mathcal{M}''$ and $\mathcal{M}'\cup\mathcal{M}''$
are pseudo $d$-manifolds, according to the previous lemma.

For {\em (ii)}, assume $H\neq \mathcal{M}_0$.
Let $x$ be a vertex in $A\setminus C$ and $S$ be a $d$-simplex that contains it.
By Fact~\ref{fact:complete_graph}, all the points of $S$ are neighbors of $x$ in the skeleton graph.
The neighbors of $x$ are all in $A$, since there is no edge between $A\setminus C$ and $C\setminus A$.
Therefore, $S$ is in $\mathcal{M}'$ but not in $\mathcal{B}\cup\mathcal{M}''$.
By symmetry of $\mathcal{M}'$ and $\mathcal{M}''$, there is also a $d$-simplex which is not in $\mathcal{M}'$.

Let us now turn to prove {\em (iii)}.
Let $S$ be an elementary $d$-complex in $\mathcal{M}$.
As $B$ separates $G_\mathcal{M}[\mathcal{M}_0\setminus H]$ into the two components
$M'$ and $M''$, it is not possible that $S$ contains elements from both $M'$
and from $M''$, as from Fact~\ref{fact:complete_graph} we know that
$G_S$ is a complete graph. So, either all the vertices of $S$ are in $B$,
or in $B\cup M'$, or they are in $B\cup M''$.
This proves that $S$ belongs to $\mathcal{B}\cup\mathcal{M}'\cup\mathcal{M}''$.
Therefore $\mathcal{M}\subseteq\mathcal{B}\cup\mathcal{M}'\cup\mathcal{M}''$.
The converse inclusion clearly holds, which implies that it is in
fact an equality. Moreover, from their definitions,
the simplicial complexes $\mathcal{B}$, $\mathcal{M}'$ and $\mathcal{M}''$
have no $d$-simplex in common. Then using the first point, two applications
of Lemma~\ref{lemma:splitting} allow us to deduce the announced equality
in {\em (iii)}.

For {\em (iv)}, let $S$ be a $(d-1)$-simplex in $\boundary[\mathcal{M}']$. It is not in $\boundary[\mathcal{M}]$ if
and only if it belongs to two $d$-simplices of $\mathcal{M}$. We will prove that
if $S$ belongs to two $d$-simplices of $\mathcal{M}$, then its $0$-faces must all lie in $B$.
Assume that $S$ is a $(d-1)$-face common to two $d$-simplices $T_1$ and $T_2$. We can assume
without loss of generality that $T_1$ belongs to $\mathcal{M}'$. But $T_2$ can not be in $\mathcal{M}'$ as $S$
is in the boundary of $\mathcal{M}'$. So, either $T_2$ is in $\mathcal{B}$, or it is in $\mathcal{M}''$.
In the first case, it immediately follows that $S$ has all its $0$-faces in $B$, as it is a face
of a $d$-simplex whose $0$-faces all lie in $B$. In the second case, again, the only possibility is
that $S$ has all its $0$-faces in $B$, as else a vertex in $M'$ and a vertex in $M''$ would be
neighbors. This proves the third point for $\boundary[\mathcal{M}']$.
The proof is the same for $\boundary[\mathcal{M}'']$.
\end{Proof}

We are now ready to state Algorithm~\ref{algo:nonoriented_sperner}
and Algorithm~\ref{algo:oriented_sperner} which solve
respectively {\bf $d$-SPM} and {\bf $d$-OSPM}
when the labels of the $0$-faces of $\partial\mathcal{M}$ are also known.
We next give the result which states the correctness of our algorithms and specifies their complexities.

\begin{Lemma}
\label{lem:algocorrect}
If $\mathcal{M}$ and $S$ satisfy the promises of the respective Sperner problems, then
Algorithms~\ref{algo:nonoriented_sperner} and~\ref{algo:oriented_sperner}
return a solution and use at most $s(G_\mathcal{M}[\mathcal{M}_0\setminus H])$ queries.
\end{Lemma}

\begin{algorithm}[H]
\caption{Main routine for solving {\bf $d$-SPM}.}
\begin{algorithmic}
\vskip 1mm
\STATE {\bf Input:} A pseudo $d$-manifold $\mathcal{M}$, $S\in\mathcal{M}_d$, a set $H\supseteq(\partial\mathcal{M})_0$ together with the labels of its elements.
\vskip 1mm
\hrule
\vskip 1mm
\STATE Let $(A,C)\prec G_\mathcal{M}[\mathcal{M}_0\setminus H]$ be a best separation, and $B=H\cup(A\cap C)$.
\STATE Let the complexes $\mathcal{B}$, $\mathcal{M}'$ and $\mathcal{M}''$ be defined as in Theorem~\ref{th:splitting}.
\STATE Query the labels of the vertices in $A\cap C$.
\IF{$\mathcal{B}$ contains a fully labeled elementary $d$-complex}
  \STATE Return the corresponding oriented $d$-simplex.
\ENDIF
\STATE Evaluate $N_{d-1}[\widehat{\boundary{\mathcal{B}}}]$, $N_{d-1}[\widehat{\boundary{\mathcal{M}'}}]$
       and $N_{d-1}[\widehat{\boundary{\mathcal{M}''}}]$.
\IF{$N_{d-1}[\widehat{\boundary{K}}]=1$ for $K\in\{\mathcal{B},\mathcal{M}',\mathcal{M}''\}$}
  \STATE Iterate on $K$, any $d$-simplex $S\in K$, and $B$ with the labels of its elements.
\ELSE
  \STATE Iterate on $K\in\{\mathcal{B},\mathcal{M}',\mathcal{M}''\}$
         containing $S$, $S$ and $B$ with the labels of its elements.
\ENDIF
\end{algorithmic}
\label{algo:nonoriented_sperner}
\end{algorithm}

\begin{algorithm}[H]
\caption{Main routine for solving {\bf $d$-OSPM}.}
\begin{algorithmic}
\vskip 1mm
\STATE {\bf Input:} A pseudo $d$-manifold $\mathcal{M}$, $S\in\mathcal{M}_d$, a set $H\supseteq(\partial\mathcal{M})_0$ together with the labels of its elements.
\vskip 1mm
\hrule
\vskip 1mm
\STATE Let $(A,C)\prec G_\mathcal{M}[\mathcal{M}_0\setminus H]$ be a best separation, and $B=H\cup(A\cap C)$.
\STATE Let the complexes $\mathcal{B}$, $\mathcal{M}'$ and $\mathcal{M}''$ be defined as in Theorem~\ref{th:splitting}.
\STATE Query the labels of the vertices in $A\cap C$.
\IF{$\mathcal{B}$ contains a fully labeled elementary $d$-complex}
  \STATE Return the corresponding oriented $d$-simplex.
\ENDIF
\STATE Evaluate $N_{d-1}[\widehat{\boundary{\mathcal{B}}}]$, $N_{d-1}[\widehat{\boundary{\mathcal{M}'}}]$
       and $N_{d-1}[\widehat{\boundary{\mathcal{M}''}}]$.
\IF{$(-1)^d N_{d-1}[\widehat{\boundary{K}}]<0$ on $K\in\{\mathcal{B},\mathcal{M}',\mathcal{M}''\}$}
  \STATE Iterate the algorithm on $K$, any $d$-simplex $S\in K$, and $B$ with the labels of its elements.
\ELSE
  \STATE Iterate the algorithm on $K\in\{\mathcal{B},\mathcal{M}',\mathcal{M}''\}$
         containing $S$, $S$ and $B$ with the labels of its elements.
\ENDIF
\end{algorithmic}
\label{algo:oriented_sperner}
\end{algorithm}

\begin{Proof}
We will prove the two claims for Algorithm~\ref{algo:nonoriented_sperner} by induction,
the proofs for Algorithm~\ref{algo:oriented_sperner} are similar.

We start by proving the correctness.
First observe that there is always enough information for the evaluations of
the flows. Indeed by {\em (iv)} of Theorem~\ref{th:splitting}, all the
$0$-faces of $\boundary[\mathcal{B}]$, $\boundary[\mathcal{M}']$ and
$\boundary[\mathcal{M}'']$ are in $\boundary[\mathcal{M}]\cup B$. The labels of the
$0$-faces of $\mathcal{M}$ are given as an input, and the labels of $B$
are queried right before the flow evaluations.

Let us now consider an input that satisfies the promise of {\bf $d$-SPM}, and
where the number of $d$-simplices in $\mathcal{M}$ is $n$.
If $n=1$, then $\mathcal{M}=\mathcal{B}$ is an elementary $d$-complex, and
by the promise, it is fully labeled. Therefore, the output of the algorithm is correct.
When $n>1$, we will prove that the recursive call will be made on an input which also
satisfies the promise, and where the number of $d$-simplices in the pseudo $d$-manifold is
less than $n$.

From Theorem~\ref{th:sperner_general} and {\em (iii)} of Theorem~\ref{th:splitting},
we have $N_{d-1}[\widehat{\boundary[\mathcal{M}]}]=N_{d-1}[\widehat{\boundary[\mathcal{B}]}]
+N_{d-1}[\widehat{\boundary[\mathcal{M}']}]+N_{d-1}[\widehat{\boundary[\mathcal{M}'']}]$.
In case {\em a)} of the promise, this sum is equal to $1$, and therefore there exists
a $K\in\{\mathcal{B},\mathcal{M}',\mathcal{M}''\}$ for which $N_{d-1}[\widehat{\boundary{K}}]=1$.
In case {\em b)} of the promise, either there exists
a $K\in\{\mathcal{B},\mathcal{M}',\mathcal{M}''\}$ for which $N_{d-1}[\widehat{\boundary{K}}]=1$,
or there exists a $K\in\{\mathcal{B},\mathcal{M}',\mathcal{M}''\}$ for which
$N_{d-1}[\widehat{\boundary{K}}]=0$ and $S\in K$.
In both cases, the number of $d$-simplices in $K$ is less than the number of $d$-simplices of $\mathcal{M}$
because of {\em (ii)} of Theorem~\ref{th:splitting}.

Let us now prove the bound on the complexity.
For every pseudo $d$-manifold $\mathcal{M}$, denote by $T(\mathcal{M},H)$ the number of queries
made by the algorithm on $\mathcal{M}$ with the set $H$ of labels.
Each recursive call ends in three possible ways: \\
\begin{tabular}{rl}
& 1) it stops after the first test if $\mathcal{M}$ is a fully labeled elementary $d$-complex, \\
& 2) it iterates on $\mathcal{B}$, \\
& 3) or it iterates on $\mathcal{M}'$ or on $\mathcal{M}''$.
\end{tabular}

\noindent In the first case, no queries are made, as all vertices of $\mathcal{M}$ are on its boundary.
In the second case, $|A\cap C|$ queries are made, as in further iterations all labels will be known.
In the third case, the number of queries is at most
$|A\cap C|+\max(T({\mathcal{M}'},B),T({\mathcal{M}''}, B))$.
Thus, we get $T(\mathcal{M},H)\leq |A\cap C|+\max(T({\mathcal{M}'},B),T({\mathcal{M}''},B))$.

We now prove that $T(\mathcal{M},H)\leq s(G_\mathcal{M}[\mathcal{M}_0\setminus H])$ for
every pseudo $d$-manifold $\mathcal{M}$ and set of vertices $H\subseteq\mathcal{M}_0$.
If $\mathcal{M}$ is an elementary $d$-complex,
then the algorithm does not make any query, and therefore $T(\mathcal{M},H)=0$,
and the statement is trivial.

Let now $\mathcal{M}$ be a pseudo $d$-manifold that is not an elementary $d$-complex.
We have $G_{\mathcal{M}'}[\mathcal{M}'_0\setminus B]\subseteq G_\mathcal{M}[M']=G_\mathcal{M}[A\setminus C]$ and
$G_{\mathcal{M}''}[\mathcal{M}_0''\setminus B]\subseteq G_\mathcal{M}[M'']=G_\mathcal{M}[C\setminus A]$, for a
best separation $(A,C)\prec G_\mathcal{M}[\mathcal{M}_0\setminus H]$. Using the induction
hypothesis, we get $T({\mathcal{M}'},B)\leq s(G_{\mathcal{M}'}[\mathcal{M}'_0\setminus B])$ and
$T({\mathcal{M}''},B)\leq s(G_{\mathcal{M}''}[\mathcal{M}''_0\setminus B])$. 
Since $s(G')\leq s(G)$ if $G'$ is a subgraph of $G$, we get
$T({\mathcal{M}'},B)\leq s(G_\mathcal{M}[A\setminus C])$
and $T({\mathcal{M}''},B)\leq s(G_\mathcal{M}[C\setminus A])$.
As $(A,C)$ is a best separation of $G_\mathcal{M}[\mathcal{M}_0\setminus H]$, this proves the inequality
$T(\mathcal{M},H)\leq s(G_\mathcal{M}[\mathcal{M}_0\setminus H])$.
\end{Proof}

\begin{Theorem}
\label{th:abstract_complexity}
$\DQC(\text{\bf $d$-SPM})=O(s(G_\mathcal{M}[\mathcal{M}_0
  \setminus(\boundary[\mathcal{M}])_0]))+|(\partial\mathcal{M})_0|$
and \\
$\,\DQC(\text{\bf $d$-OSPM})=O(s(G_\mathcal{M}[\mathcal{M}_0
  \setminus(\boundary[\mathcal{M}])_0]))+|(\partial\mathcal{M})_0|$.
\end{Theorem}

\begin{Proof}
The algorithms consist in querying the labels of the vertices of $\partial\mathcal{M}$ and
then running respectively Algorithm~\ref{algo:nonoriented_sperner} or Algorithm~2
with the initial choice $H=(\boundary[\mathcal{M}])_0$.
\end{Proof}

To bound the complexity of our algorithms we need an upper-bound on the iterated separator number of the skeleton graph.
The following theorem gives, for any graph, an upper bound on the size of a balancing separator,
whose deletion leaves the graph with two roughly equal size components.
The bound depends on the genus and the number of vertices of the graph.
\begin{Theorem}[Gilbert, Hutchinson, Tarjan~\cite{separator_genus}]
\label{th:separation_genus}
A graph of genus $g$ with $n$ vertices has a set of at most $6\sqrt{g\cdot n}+2\sqrt{2n}+1$ vertices
whose removal leaves no component with more than $2n/3$ vertices.
\end{Theorem}

For our purposes we can immediately derive an upper bound on the iterated separation number.

\begin{Corollary}
\label{cor:genus}
For graphs $G=(V,E)$ of size $n$ and genus $g$ we have 
$s(G)\leq \lambda(6\sqrt{g\cdot n}+2\sqrt{2n})+\log_{3/2} n$,
where $\lambda$ is solution of $\lambda=1+\lambda\sqrt{2/3}$.
\end{Corollary}

\begin{Proof}
Let us prove this fact by induction over $n$.
It obviously holds for $n=1$.
Assume now that $n>1$.
Theorem~\ref{th:separation_genus} shows that there 
exist three pairwise disjoint sets $S_1, S_2$ and $S_3$ such that $V=S_1\cup S_2 \cup S_3$,
$|S_2|\leq 6\sqrt{g\cdot n}+2\sqrt{2n}+1$ and $|S_1|,|S_3|\leq 2n/3$.
If we let $A=S_1\cup S_2$ and $C=S_2\cup S_3$, then $(A,C)\prec G$ and $A\cap C=S_2$.
The construction
implies that $|A\setminus C|,|C\setminus A|\leq 2n/3$.
Using the induction hypothesis, we get
\begin{align*}
s(G) & \leq |A\cap C|+\max(s(G[A\setminus C]),s(G[C\setminus A])) \\
  & \leq 6\sqrt{g\cdot n}+2\sqrt{2n}+1+\lambda(6\sqrt{g\cdot2n/3}+2\sqrt{2\cdot 2n/3})+\log_{3/2}(2n/3) \\
  & \leq \lambda(6\sqrt{g\cdot n}+2\sqrt{2n})+\log_{3/2}n.
\end{align*}
\end{Proof}

In general, there is no immediate relationship between the genus of a pseudo $d$-manifold
and the genus of its skeleton graph. However, if the pseudo $d$-manifold $\mathcal{M}$ is
a triangulated oriented surface, then the genus of the graph is equal to the genus of $\mathcal{M}$.
Used in conjunction with Corollary~\ref{cor:genus}, Theorem~\ref{th:abstract_complexity} gives
an effective upper bound for pseudo $d$-manifolds.

\begin{Corollary}
Let $\mathcal{M}$ be a pseudo $d$-manifold such that 
$G_\mathcal{M}$ is of size $n$ and of genus $g$.
Then, $ \DQC(\text{\bf $d$-SPM})=O(\sqrt{g})\cdot\sqrt{n}+|(\partial\mathcal{M})_0|$ and
$\DQC(\text{\bf $d$-OSPM})=O(\sqrt{g})\cdot\sqrt{n}+|(\partial\mathcal{M})_0|$.
\end{Corollary}

Since the skeleton graph of the underlying pseudo $2$-manifold of {\bf REGULAR $2$-SPM} is planar,
it has genus $0$. Thus we get:

\begin{Theorem}
\label{th:sperner_algorithm}
$\DQC(\text{\bf REGULAR $2$-SPM})=O(\sqrt{n})$.
\end{Theorem}

In the next section, we show nontrivial lower bounds on the randomized and the quantum query complexity of the
{\bf REGULAR $2$-SPM} problem. Observe that for some general instances of the {\bf $2$-SPM}
over the same pseudo $2$-manifold we can easily derive exact lower bounds from the known
complexity of Grover's search problem~\cite{Grover}.
For example, if a labeling is $2$ everywhere, except on two consecutive vertices
on the boundary where it takes respectively the values $0$ and $1$, then finding a fully labeled $2$-simplex
is of the same complexity as finding a distinguished element on the boundary.

\section{Lower bounds for {\bf REGULAR $2$-SPM}}
\label{section:bounds}
We denote by {\bf UNIQUE-SPERNER} all those instances of {\bf REGULAR $2$-SPM} for which there exists
a unique fully labeled triangle.
There exist several equivalent adversary methods for proving quantum lower bounds in the query model~\cite{Mario_Robert}.
Here, we will use the weighted adversary method~\cite{Aaronson_min_search,Ambainis2,Laplante_Magniez}.

\begin{Theorem}
\label{th:weights}
Let $\Sigma$ be a finite set, let $n\geq 1$ be an integer, and let $S\subseteq\Sigma^n$ and
$S'$ be sets. Let $f:S\toset S'$. Let $\Gamma$ be an arbitrary $S\times S$ nonnegative symmetric matrix
that satisfies $\Gamma[x,y]=0$ whenever $f(x)=f(y)$. For $1\leq k\leq n$, let $\Gamma_k$ be the matrix
such that $\Gamma_k[x,y]=0$ if $x_k=y_k$, and $\Gamma_k[x,y]=\Gamma[x,y]$ otherwise.
For all $S\times S$ matrix M and $x\in S$, let $\sigma(M,x)=\sum_{y\in S} M[x,y]$. Then
$$ \QQC(f)=\Omega\left(\min_{\Gamma[x,y]\neq 0,x_k\neq y_k}\sqrt{\frac{\sigma(\Gamma,x)\sigma(\Gamma,y)}
  {\sigma(\Gamma_k,x)\sigma(\Gamma_k,y)}}\right), $$
$$ \RQC(f)=\Omega\left(\min_{\Gamma[x,y]\neq 0,x_k\neq y_k}\max\left(\frac{\sigma(\Gamma,x)}{\sigma(\Gamma_k,x)},
   \frac{\sigma(\Gamma,y)}{\sigma(\Gamma_k,y)}\right)\right). $$
\end{Theorem}

For the lower bound we will consider specific instances of {\bf REGULAR $2$-SPM}.
For that, we need a few definitions.
For any binary sequence $b$, let $|b|$ denote the length of the sequence $b$, and
for $i=0,1$ 
let $w_i(b)$ be the number of bits $i$ in $b$. For
$0\leq t \leq |b|$, let $b^t=b_1\ldots b_t$ denote the prefix of length $t$ of $b$.

The instances of {\bf REGULAR $2$-SPM} we will consider are those whose oracle inputs $C_b$ are
induced by binary sequences $b=b_1 \ldots  b_{m-2}$ of length $m-2$ as follows:
{
$$
C_b(i,j)=\begin{cases}
1 & \text{ if } j=0 \text{ and } i\neq 0, \\
2 & \text{ if } i=0 \text{ and } j\neq m, \\
0 & \text{ if } i+j=m \text{ and } j\neq 0, \\
1 & \text{ if there exists }  0 \leq t \leq m-2   \text{ with }  (i,j)=(w_0(b^t)+1,w_1(b^t)), \\
2 & \text{ if there exists }  0 \leq t \leq m-2   \text{ with }   (i,j)=(w_0(b^t),w_1(b^t)+1), \\
0 & \text{ otherwise.}
\end{cases}
$$
}

\begin{figure}[t]
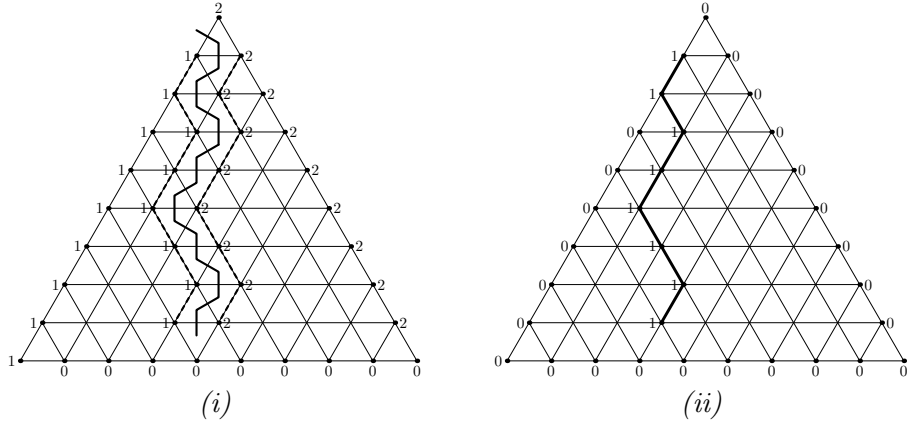

\begin{center}
\begin{tabular}{ccc}
\epsfig{file=Query.2,width=5.5cm} & \hphantom{ } & \epsfig{file=Query.3,width=5.5cm} \\
{\em (i)} & & {\em (ii)}
\end{tabular}
\end{center}
\caption{
In the coordinates system of the Figure, the point $(0,0)$ is the highest corner of the triangles,
the $x$ coordinates increase by going down and left, and the $y$ coordinates increase by going
down and right.
On sub-figure {\em (i)}, the labeling $C_b$ corresponds to the binary sequence $b=0100110$.
On sub-figure {\em (ii)}, the labeling $O_b$ corresponds to the same sequence $b$.
The unmarked vertices are all labeled $0$.}
\label{fig:walk_example}
\end{figure}

Notice that the first and fourth~({\em resp.} second and fifth) conditions can be simultaneously
satisfied, but the labeling definition is consistent.
Also observe that, for any $b$, there is a unique fully labeled triangle, whose coordinates
are $\{(w_0(b)+1,w_1(b)), (w_0(b),w_1(b)+1), (w_0(b)+1,w_1(b)+1)\}$. 
Therefore $C_b$ is an instance of {\bf UNIQUE-SPERNER}. We illustrate an instance of $C_b$
in Figure~\ref{fig:walk_example}.

It turns out that technically it will be easier to prove the lower bound for a problem which is closely related
to the above instances of {\bf REGULAR $2$-SPM}, that we call {\bf SNAKE}.
Recall that
$V_m =\{(i,j)\in\Nm^2 \,|\,  0 \leq i+j\leq m\}$. 
For every binary sequence $b= b_1 \ldots  b_{m-2}$, we denote by $O_b$ the function $V_m\toset\{0,1\}$ defined
for $p\in V_m$ by
{
$$
O_b(p)=\begin{cases}
1 & \text{ if there exists }  0 \leq t \leq m-2   \text{ with } (i,j)=(w_0(b^t)+1,w_1(b^t)), \\
0 & \text{ otherwise.}
\end{cases}
$$
}
See again Figure~\ref{fig:walk_example} for an example.

{
\vskip 2mm
\noindent {\bf SNAKE} \\
\begin{tabular}{ll}
  {\it Input:} & $n =  {m \choose 2}$ for some integer $m.$\\
  {\it Oracle input:} & a function $f:V_m\toset\{0,1\}$. \\
  {\it Promise:} & there exists a binary sequence $b= b_1 \ldots  b_{m-2}$ such that $f=O_b$. \\
  {\it Output:} & $(w_0(b),w_1(b))$.\\
\end{tabular}
\vskip 2mm
}

We recall here the definition of~\cite{Miklos_Mario} of $c$-query reducibility between black-box problems, which
we will use to prove our lower bound.
\begin{Definition}
For an integer $c > 0,$
a functional oracle problem
$A : S_1 \rightarrow T_1$ with $S_1 \subseteq \Sigma_1^{n}$
is $c$-{\em query reducible} to a functional oracle problem
$B : S_2 \rightarrow T_2$ with $S_2 \subseteq \Sigma_2^{n'}$
if the following two conditions hold:
\begin{enumerate}[(i)]
\item $\exists \alpha : S_1 \rightarrow S_2, \quad \exists \beta:T_2\toset T_1,$ such that
  $\forall x \in S_1, \  A(x)=\beta(B(\alpha(x)))$,  \label{cond1}
\item $\exists \gamma_1, \ldots, \gamma_c : \{1, \ldots , n'\} 
\rightarrow  \{1, \ldots , n\}$
  and $\gamma : \{1, \ldots , n'\}
  \times \Sigma_1^c \rightarrow \Sigma_2$ such that
  $\forall  x \in S_1,\, k\in \{1, \ldots , n'\}, 
  \quad\alpha(x)(k) = \gamma(k, x_{\gamma_1(k)}, \ldots, x_{\gamma_c(k)})$. \label{cond2}
\end{enumerate}
\end{Definition}
                                                                                                               
\begin{Lemma}[\cite{Miklos_Mario}]
\label{lemma:reduction}
If $A$ is $c$-query reducible to $B$ then
$\QQC(B) \geq \QQC(A)/2c$, and $\RQC(B) \geq \RQC(A)/c$.
\end{Lemma}

\begin{Lemma}
\label{lemma:sperner_reduction}
{\bf SNAKE} is 3-query reducible to {\bf UNIQUE-SPERNER}.
\end{Lemma}

\begin{Proof}
We define the oracle transformations as $\alpha(O_b)=C_b$, and
$$\beta(\{(i_1,j_1),(i_2,j_2),(i_3,j_3)\})=(\min\{i_1,i_2,i_3\},\min\{j_1,j_2,j_3\}).$$
Obviously, $\alpha$ and $\beta$ satisfy the first condition of the definition.

We now turn to the simulation of an oracle for {\bf UNIQUE-SPERNER} by an oracle for {\bf SNAKE}.
If the query concerns a point $(i,j)$ on the boundary, the answer is independent from the oracle,
and is given according to the definition of $C_b$, for any $b$. Otherwise, the simulator will
query the point and its left and right neighbors in the sense of the Figure~\ref{fig:walk_example},
from which the value of $C_b$ can be easily determined.
Formally, for such a point $(i,j)$, let the functions $\gamma$, $\gamma_1$, $\gamma_2$ and $\gamma_3$ be defined as
\begin{equation*}
\gamma((i,j),a_1,a_2,a_3) =\begin{cases}
1 & \text{ if } (a_1,a_2,a_3)=(0,1,0), \\
2 & \text{ if } (a_1,a_2,a_3)=(1,0,0), \\
0 & \text{otherwise,}
\end{cases}
\end{equation*}
and
$\gamma_1(i,j) = (i+1,j-1)$,
$\gamma_2(i,j) = (i,j)$,
$\gamma_3(i,j) = (i-1,j+1)$.
\end{Proof}

\begin{Lemma}
\label{lemma:snakes}
The query complexity of {\bf SNAKE} f satisfies
$$\RQC(\text{\bf SNAKE}) =\Omega(\sqrt[4]{n}) \text{ and }
\QQC(\text{\bf SNAKE}) =\Omega(\sqrt[8]{n}).$$
\end{Lemma}

\begin{Proof}
We give now the definition of the adversary matrix $\Gamma$ which will be a $2^{m-2}\times 2^{m-2}$
symmetric matrix, whose rows and columns will be indexed by the labelings $O_b$, when $b\in\{0,1\}^{m-2}$.
For the sake of simplicity, we will only use binary sequences to denote rows and columns, instead of the
induced labelings. For two binary sequences $b$ and $b'$, we denote by $b\wedge b'$ their longest common prefix.
Then let
$$
\Gamma[b,b']=\begin{cases}
0 & \text{ if } w_0(b)=w_0(b'), \\
2^{|b\wedge b'|} & \text{ otherwise.}
\end{cases}
$$
For a given binary sequence $b$, there are $2^{m-2-(d+1)}$ sequences that have longest common
prefix of length $d$ with $b$. Out of them, $m-3-d \choose w_{b_{d+1}}(b)-w_{b_{d+1}}(b^d)$ will give the same
output as $b$. Therefore,
\begin{align*}
\sigma(\Gamma,b) &=\sum_{d=0}^{m-4} 2^d\left[2^{m-3-d}-{m-3-d \choose w_{b_{d+1}}(b)-w_{b_{d+1}}(b^d)}\right] \\
  & \geq (m-3)2^{m-3}-\sum_{d=0}^{m-4} 2^d{m-3-d \choose \lfloor (m-3-d)/2\rfloor} \\
  & \geq (m-3)2^{m-3}-\left(\sum_{d=0}^{m-4} \frac{2^{m-3}}{\sqrt{m-3-d}}\right) \\
  & \geq (m-3)2^{m-3}-O(\sqrt{m} 2^m)=\Omega(m 2^m).
\end{align*}

We now turn to bound from above $\sigma(\Gamma_p,b)$ and $\sigma(\Gamma_p,b')$ when $\Gamma[b,b']\neq 0$
and $O_b(p)\neq O_{b'}(p)$. Let us now fix a point $p=(i,j)$ with $1\leq i+j\leq m-1$,
and two sequences $b$ and $b'$, such that $O_b(p)\neq O_{b'}(p)$. We assume that $O_b(p)=0$ and $O_{b'}(p)=1$.
We trivially upper bound $\sigma(\Gamma_p,b')$ by $\sigma(\Gamma,b')=O(m2^m)$.

We will now upper-bound $\sigma(\Gamma_p,b)$.
Set $h=i+j$. If a sequence $b''$ is such that $O_{b''}(p)=1$,
then the length of its longest common prefix with $b$ is at most $h-2$.
We regroup these sequences according to the value $|b\wedge b''|$. The number of sequences $b''$ for
which $|b\wedge b''|=d$ and $O_{b''}(p)=1$ is at most ${h-1-(d+1) \choose \lfloor h-1-(d+1))/2\rfloor}2^{m-h-1}$.
Therefore we can bound
$\sigma(\Gamma_p,b)$ as
\begin{align*}
\sigma(\Gamma_p,b) & \leq \sum_{d=0}^{h-2} 2^d\cdot {h-d-2 \choose \lfloor(h-d-2)/2\rfloor}2^{m-h-1} \\
 & \leq \sum_{d=0}^{h-2} \left[2^{-(h-d-2)}\cdot {h-d-2 \choose \lfloor(h-d-2)/2\rfloor}\right]2^{m-3} =O(\sqrt{m}2^m).
\end{align*}

By Theorem~\ref{th:weights} we conclude that
\begin{align*}
\RQC(\text{\bf SNAKE}) & =\Omega\left(\max\left(\frac{m2^m}{m2^m},\frac{m2^m}{\sqrt{m}2^m}\right)\right)=\Omega(\sqrt[4]{n}), \\
\QQC(\text{\bf SNAKE}) & =\Omega\left(\frac{m2^m}{\sqrt{m2^m\cdot \sqrt{m}2^m}}\right)=\Omega(\sqrt[8]{n}).
\end{align*}
\end{Proof}

\begin{Theorem}
\label{th:blackbox}
The query complexity of {\bf REGULAR $2$-SPM} 
satisfies
$$
\RQC(\text{\bf REGULAR $2$-SPM})=\Omega(\sqrt[4]{n}) \text{ and }
\QQC(\text{\bf REGULAR $2$-SPM})=\Omega(\sqrt[8]{n}).
$$
\end{Theorem}

\begin{Proof}
By Lemma~\ref{lemma:reduction} and~\ref{lemma:sperner_reduction}, the lower bounds of Lemma~\ref{lemma:snakes} for {\bf SNAKE}
also apply to {\bf REGULAR $2$-SPM}.
\end{Proof}

%%--------------------------------------------------------------------------------------

%%----------------------------------------------------------------------------------------


\begin{thebibliography}{10}
\setlength{\itemsep}{-2pt}
\bibitem{Aar2}
S. Aaronson.
\newblock Quantum lower bound for the collision problem.
\newblock In {\em 34th  STOC}, pp. 635--642, 2002.

\bibitem{Aaronson_min_search}
S. Aaronson.
\newblock Lower bounds for local search by quantum arguments.
\newblock In {\em 36th STOC}, pp. 465--474, 2004.

\bibitem{Ambainis2}
A. Ambainis.
\newblock Polynomial degree vs. quantum query complexity.
\newblock In {\em 44th FOCS}, pp. 230--239, 2003.

\bibitem{BBC}
R. Beals, H. Buhrman, R. Cleve, M. Mosca and R. de Wolf.
\newblock Quantum lower bounds by polynomials,
\newblock {\em J. of the ACM} (48):4, pp. 778--797, 2001.

\bibitem{Beame}
P. Beame, S. Cook, J. Edmonds, R. Impagliazzo and T.
  Pitassi.
\newblock The relative complexity of {NP} search problems.
\newblock {\em J. Comput. System Sci.}, 57(1):3--19, 1998.

\bibitem{Grover}
C. Bennett, E. Bernstein, G. Brassard and U. Vazirani.
\newblock Strength and weaknesses of quantum computing.
\newblock {\em SIAM J. on Computing}, 26(5):1510--1523, 1997.

\bibitem{Bloch}
E. Bloch.
\newblock Mod 2 degree and a generalized no retraction {T}heorem.
\newblock To appear in {\em Mathematische Nachrichten}.

\bibitem{Morioka}
J. Buresh-Oppenheim and T. Morioka.
\newblock {Relativized NP search problems and propositional proof systems}.
\newblock In {\em 19th Conference on Computational Complexity}, pp.
  54--67, 2004.

\bibitem{Crescenzi}
P.~Crescenzi and R.~Silvestri.
\newblock Sperner's lemma and robust machines.
\newblock {\em Comput. Complexity}, 7(2):163--173, 1998.

\bibitem{DJ}
D. Deutsch and R. Jozsa.
\newblock Rapid solution of problems by quantum computation.
\newblock {\em Proc. of the Royal Society A}, volume 439,1985.

\bibitem{KyFan}
K. Fan.
\newblock Simplicial maps from an orientable $n$-pseudomanifold into ${S}^m$
  with the octahedral triangulation.
\newblock {\em J. Combinatorial Theory}, 2:588--602, 1967.

\bibitem{separator_genus}
J. Gilbert, J. Hutchinson and R. Tarjan.
\newblock A separator theorem for graphs of bounded genus.
\newblock {\em J. Algorithms}, 5(3):391--407, 1984.

\bibitem{Laplante_Magniez}
S. Laplante and F. Magniez.
\newblock Lower bounds for randomized and quantum query complexity using
  kolmogorov arguments.
\newblock In {\em 19th Conference on Computational Complexity}, pp.
  294--304, 2004.

\bibitem{Llewellyn_Tovey_2}
D. Llewellyn and C. Tovey.
\newblock Dividing and conquering the square.
\newblock {\em Discrete Appl. Math.}, 43(2):131--153, 1993.

\bibitem{Llewellyn_Tovey_1}
D. Llewellyn, C. Tovey and M. Trick.
\newblock Local optimization on graphs.
\newblock {\em Discrete Appl. Math.}, 23(2):157--178, 1989.

\bibitem{Megiddo_Papadimitriou}
N. Megiddo and C. Papadimitriou.
\newblock On total functions, existence theorems and computational complexity.
\newblock {\em Theoret. Comput. Sci.}, 81:317--324, 1991.

\bibitem{pap}
C. Papadimitriou.
\newblock On graph-theoretic lemmata and complexity classes. 
\newblock In {\em 31st FOCS}, pp. 794--801, 1990.

\bibitem{papadimitriou}
C. Papadimitriou.
\newblock On the complexity of the parity argument and other inefficient proofs
  of existence.
\newblock {\em J. Comput. System Sci.}, 48(3):498--532, 1994.

\bibitem{Miklos_Mario}
M. Santha and M. Szegedy.
\newblock Quantum and classical query complexities of local search are
  polynomially related.
\newblock In {\em 36th STOC}, pp. 494--501, 2004.

\bibitem{Shi}
Y. Shi.
\newblock Quantum lower bounds for the collision and the element distinctness problems.
\newblock In {\em 43rd FOCS}, pp. 513--519, 2002.

\bibitem{Sim}
D. Simon.
\newblock On the power of quantum computation.
\newblock {\em SIAM J. on Computing} (26):5, pp. 1474--1783, 1997.

\bibitem{Sperner}
E. Sperner.
\newblock Neuer Beweis f\"ur die Invarianz der Dimensionzahl und des Gebietes.
\newblock {\em Abh. Math. Sem. Hamburg Univ.} 6:265--272, 1928.

\bibitem{Mario_Robert}
R. \v{S}palek and M. Szegedy.
\newblock All quantum adversary methods are equivalent.
\newblock {\tt http://xxx.lanl.gov/abs/quant-ph/0409116}.

\bibitem{Taylor}
L. Taylor.
\newblock {S}perner's {L}emma, {B}rouwer's {F}ixed {P}oint {T}heorem, {T}he
  {F}undamental {T}heorem of {A}lgebra.
\newblock  {\tt
  http://www.cs.csubak.edu/\~{}larry/math/sperner.pdf}.
  
\bibitem{Thomason}
A. Thomason.
\newblock Hamilton cycles and uniquely edge colourable graphs.
\newblock {\em Ann. Discrete Math.} 3: 259--268, 1978.

\end{thebibliography}
\end{document}